\documentclass[twocolumn,aps,showpacs,superscriptaddress,pre]{revtex4}

\usepackage{amsmath,amssymb,graphicx}
\usepackage{algorithmic}
\usepackage{enumerate}
\usepackage{times}

\usepackage{color}

\definecolor{yblue}{rgb}{0.06, 0.3, 0.57}
\usepackage[pdftex]{hyperref}
\hypersetup{colorlinks=true,linkcolor=yblue,citecolor=yblue,urlcolor=yblue}

\newcommand{\ds}{d_{\rm s}/2}

\begin{document}

\title{Bond chaos in spin glasses revealed through thermal boundary conditions}

\author{Wenlong Wang}
\email{wenlong@physics.umass.edu}
\affiliation{Department of Physics, University of Massachusetts,
Amherst, Massachusetts 01003 USA}
\affiliation{Department of Physics and Astronomy, Texas A$\&$M University,
College Station, Texas 77843-4242, USA}

\author{Jonathan Machta}
\email{machta@physics.umass.edu}
\affiliation{Department of Physics, University of Massachusetts,
Amherst, Massachusetts 01003 USA}
\affiliation{Santa Fe Institute, 1399 Hyde Park Road, Santa Fe, 
New Mexico 87501, USA}

\author{Helmut G. Katzgraber}
\affiliation{Department of Physics and Astronomy, Texas A$\&$M University, 
College Station, Texas 77843-4242, USA}
\affiliation{Santa Fe Institute, 1399 Hyde Park Road, Santa Fe, 
New Mexico 87501, USA}

\begin{abstract}

Spin glasses have competing interactions that lead to a rough energy
landscape which is highly susceptible to small perturbations. These
chaotic effects strongly affect numerical simulations and, as such,
gaining a deeper understanding of chaos in spin glasses is of much
importance. The use of thermal boundary conditions is an effective
approach to study chaotic phenomena. Here, we generalize population
annealing Monte Carlo, combined with thermal boundary conditions, to
study bond chaos due to small perturbations in the spin-spin couplings
of the three-dimensional Edwards-Anderson Ising spin glass. We show that
bond and temperature-induced chaos share the same scaling exponents and
that bond chaos is stronger than temperature chaos.

\end{abstract}

\maketitle

\section{Introduction}

Chaos is a common phenomenon in nonlinear dynamical systems.
Interestingly, complex systems with quenched disorder and frustration
often display chaotic behavior in one form or another. For example, in
spin glasses, the thermodynamic state in
thermal equilibrium can change chaotically when an external parameter,
e.g., the temperature is tuned. This is also the case for small
perturbations of the couplings between the spins, as well as random
time-dependent local biases (field chaos).  The corresponding chaotic
phenomena are therefore called temperature chaos
\cite{mckay:82,kondor:89,parisi:84,fisher:86,bray:87,ritort:94,rizzo:03,rizzo:06,sasaki:03,sasaki:05,katzgraber:07,thomas:11e,fernandez:13,wang:15},
i.e., when the temperature is changed, and bond chaos
\cite{sasaki:05,katzgraber:07}, when the interactions between the spins
are changed.  Chaos is believed to be related to hysteresis phenomena,
memory and rejuvenation effects in spin glasses
\cite{fisher:91d,sales:02,silveira:04,doussal:05}, as well as the
generic computational hardness of disordered systems
\cite{fernandez:13,wang:15a,martin-mayor:15}. Therefore, chaos is
related to both the equilibrium and nonequilibrium properties of
dynamical systems.  Chaos is also of paramount importance in analog
optimization machines, such as the D-Wave Systems Inc.~D-Wave 2X quantum
annealer. Given the intrinsic analog implementation of the device, small
problem misspecifications might lead to the solution of a different
Hamiltonian altogether. While temperature is well controlled in these
analog machines, precisely encoding the spin-spin interactions (or
qubit-qubit couplers) has proven to be difficult due to the limited
precision of the device.  Given the importance of bond chaos to these
novel computing paradigms \cite{zhu:16,martin-mayor:15}, in this work we
investigate bond chaos of spin glasses as a prototypical example of the
effects of changing the quenched disorder -- a problem far less studied
than temperature chaos.

One intriguing numerical result \cite{sasaki:05,katzgraber:07} is that
temperature chaos and bond chaos \cite{krzakala:05} appear to follow the
\textit{same} scaling properties, and that bond chaos is
\textit{considerably} stronger than temperature chaos. However, these
observations were left unexplained in
Refs.~\cite{sasaki:05,katzgraber:07}.  Here, we provide simple
explanations of both results within the framework of the droplet/scaling
picture \cite{fisher:86,fisher:88,fisher:88b,bray:86,mcmillan:84b} by
assuming that temperature chaos is mainly entropy driven, whereas bond
chaos is mainly energy driven.  Previous work have studied chaotic
effects using correlation and overlap functions between different
parameters. Here we use an alternate approach: we study bond chaos using
thermal boundary conditions \cite{wang:14,wang:15a}, in which all $2^d$
combinations of periodic and anti-periodic boundary conditions in the
$d$ directions (space dimensions) appear in the simulation with their
appropriate Boltzmann weights. In this setting, the weights of boundary
condition crossings mimic the exchange of dominance of the more abstract
pure states.  Furthermore, we generalize the population annealing (PA)
Monte Carlo \cite{hukushima:03,zhou:10,machta:10,wang:15e} algorithm to
simulate bond chaos in glassy systems.  One advantage of this approach
is that many disorder realizations up to a small perturbation can be
studied in a single simulation run, yet have enough dynamical range in
the perturbations to study the scaling properties of chaotic effects.

The paper is organized as follows. We first discuss the model,
simulation methods, and scaling properties of bond chaos in
Sec.~\ref{mm}, followed by numerical results on bond chaos compared to
previous results on temperature chaos in Sec.~\ref{results}. Concluding
remarks are stated in Sec.~\ref{cc}.

\section{Models and methods}
\label{mm}

In this section we review technical details of our study, such as the
model, numerical and analysis methods, as well as simulation details.

\subsection{Model}

We simulate the three-dimensional Edwards-Anderson (EA) Ising spin glass
represented by the Hamiltonian
\begin{equation}
H = -\sum\limits_{\langle ij \rangle} J_{ij} S_i S_j,
\end{equation}
where $S_i \in \{\pm 1\}$ are Ising spins. The sum $\langle ij \rangle$
is over the nearest neighbor sites in a cubic lattice with $N = L^3$
sites.  The coupling $J_{ij}$ between spins $S_i$ and $S_j$ is chosen
from a Gaussian distribution with mean $0$ and variance $1$.  We refer
to each disorder realization, ${\mathcal J}=\{ J_{ij} \}$, as a
``sample.''

\subsection{Generalized population annealing Monte Carlo}

We use population annealing to obtain equilibrium states at a low
temperature $T=1/\beta$ for a fixed disorder realization ${\mathcal J}$
and then obtain equilibrium states as the couplings are continuously
transformed from ${\mathcal J}$ to ${\mathcal J}'$ while $T$ is kept
fixed.  Reference \cite{wang:15e} provides a detailed description of
population annealing Monte Carlo (hereafter referred to as PA).  First,
we briefly review the ``standard'' population annealing Monte Carlo
method for maintaining thermal equilibrium as the temperature is
lowered following a given annealing schedule.

Suppose we have an ensemble or population of $R$ independent replicas of
the system chosen from the Gibbs distribution for sample ${\mathcal J}$
at inverse temperature $\beta = 1/T$.  We would like to create a
population chosen from the Gibbs distribution for sample ${\mathcal J}$
at inverse temperature $\beta'$, with $\beta' > \beta$. To achieve this
goal, we {\it resample} the population, making copies of low-energy
replicas and removing high-energy replicas from the population. If
replica $i$ has energy $E_i$, then the expected number of copies
$\tau_i$ to make of replica $i$ is
\begin{equation}
\tau_i=\frac{1}{Q}e^{-(\beta'-\beta) E_i} 
\;\;\; {\rm with} \;\;\;
Q=\frac{1}{R_0}\sum\limits_i e^{-(\beta'-\beta) E_i} . 
\label{eq:tch}
\end{equation}
Here, $R_0$ is the expected population size. Note that $\tau_i$ is
proportional to the ratio of the Boltzmann factors at the two
temperatures for a system with energy $E_i$, normalized such that the
sum of the $\tau$'s is equal to the target population size $R_0$.

Now suppose we have a population in equilibrium at inverse temperature
$\beta$ with interactions ${\mathcal J}$ and we would like to transform
to a new population at the same temperature $\beta$ but with {\em
different} couplings ${\mathcal J}'$. For such a procedure, the
resampling now requires that 
\begin{equation}
\tau_i=\frac{1}{Q}e^{-\beta(E_i'-E_i)} 
\;\;\; {\rm with} \;\;\;
Q=\frac{1}{R_0}\sum\limits_i e^{-\beta(E_i'-E_i)} .
\label{eq:jch}
\end{equation}
In Eq.~\eqref{eq:jch}, $E_i'$ and $E_i$ are the energy of replica $i$
with bonds ${\mathcal J}'$ and ${\mathcal J}$, respectively, with the
spins in each replica held fixed.

One can summarize Eqs.~\eqref{eq:tch} and \eqref{eq:jch} using reduced
Hamiltonians $\mathcal{H} = \beta H$.  For two sets of close (similar)
distributions on the same state space with Boltzmann factors of
$\exp[-\mathcal{H}_i]$ and $\exp[-\mathcal{H}_i']$ for replica $i$, when
transforming a population of replicas from $\mathcal{H}$ to
$\mathcal{H'}$,
\begin{equation}
\tau_i=\frac{1}{Q}e^{-(\mathcal{H}_i'-\mathcal{H}_i)}
\;\;\; {\rm with} \;\;\;
Q=\frac{1}{R_0}\sum\limits_i e^{-(\mathcal{H}_i'-\mathcal{H}_i)}.
\label{eq:gch}
\end{equation}
Within this framework, one can perturb the system either by a change in
temperature, by a change in the spin-spin interactions, or both using
PA. Finally, we note that the free-energy difference can be generalized,
too, using the free-energy perturbation method as $-\beta' F' = -\beta F
+ \langle \exp[-(\mathcal{H'}-\mathcal{H})] \rangle_{\mathcal{H}}$,
although the free energy is not needed in this work.

One advantage of PA over other methods such as parallel tempering Monte
Carlo \cite{hukushima:96} is the ease of simulating thermal boundary
conditions. In thermal boundary conditions, each of the $2^d$ choices of
periodic or anti-periodic boundary conditions in each of the $d$
Cartesian directions appears in the ensemble with its correct
statistical weight, given by $\exp(-\beta F_i)$, where $F_i$ is the free
energy of the system in boundary condition $i$.  As described in
Ref.~\cite{wang:15e}, thermal boundary conditions can be simulated in PA
by initiating the population at $\beta=0$ with an equal fraction $1/2^d$
of the population is in each of the $2^d$ boundary conditions.
Thereafter, as the temperature or the bond configuration is modified,
resampling changes the relative fraction of each boundary condition. At
each value of $\beta$ and $\mathcal{J}$ in the annealing schedule, let
$p_i$ be the fraction of boundary condition $i$ in the population.  The
free energy of each boundary condition is then proportional to $-\log
p_i$. The evolution of $p_i$ with bond configuration is our main tool to
study bond chaos.

\subsection{Scaling analysis and observables}

We present the scaling analysis of temperature and bond chaos within the
droplet theory of the low-temperature phase of the the Edwards-Anderson
Ising spin glass. In this theory, the low-lying excitations of the spin
glass are flipped compact droplets. The free energy cost to flip a
droplet of size $\ell$ at temperature $\beta$ scales as $\ell^{\theta}$
with disorder ${\mathcal J}$, and the free energy cost to perturb the
bonds with $\delta {\mathcal J}$ for the droplet and the flipped droplet
is $\Delta F_1$ and $\Delta F_2$, respectively.  Then the free-energy
cost to flip the droplet at ${\mathcal J}'= {\mathcal J} + \delta
{\mathcal J}$ is $\ell^{\theta}+\Delta F_2 -\Delta F_1$, with $\theta$
the stiffness exponent.  One can see that the effect of changes in the
spin-spin interactions for the last two terms is nonzero only at the
surface of the droplet due to spin-reversal symmetry.  Because $\Delta F
= \Delta E - T \Delta S$, and assuming the energy difference dominates
when we change the couplings and therefore, the last two terms scale as
$\delta {\mathcal J}\ell^{\ds}$, where $\ds$ is the fractal dimension of
the boundary of the droplet.  Putting everything together, the
free-energy cost to flip the droplet at ${\mathcal J}'$ scales as
$\ell^{\theta} - \delta {\mathcal J}\ell^{\ds}$, and therefore the
strength $\delta {\mathcal J}$ needed for bond chaos scales as
\begin{equation}
\delta {\mathcal J} \sim \frac{1}{\ell^{\zeta}},
\end{equation}
where 
\begin{equation}
\zeta=\ds-\theta .
\label{eq:relation}
\end{equation}
This simple derivation using droplet arguments suggests that bond chaos
effects should be described by the same scaling exponents ($\theta$,
$\ds$ and $\zeta$) as temperature chaos \cite{wang:15a,katzgraber:07}.

Following Refs.~\cite{wang:14,wang:15a}, $\theta$ can be calculated
using sample stiffness scaling, $\ds$ can be calculated using the
scaling of energy differences at boundary condition crossings, and
$\zeta$ is related to the scaling of the number of dominant crossings.
We briefly summarize these quantities and their scaling. 

For a sample ${\mathcal J}$ at inverse temperature $\beta$, let
$f_{{\mathcal J},\beta}=\max_i\left[p_i\right]$ be the fraction of the
population in the {\em dominant} boundary condition, i.e., the
boundary condition with the largest fraction in the population. The
sample stiffness $\lambda_{{\mathcal J},\beta}$ is defined as
\begin{equation}
\label{eq:lam}
\lambda_{{\mathcal J},\beta} = \log\frac{f_{{\mathcal J},\beta}}{1-f_{{\mathcal J},\beta}},
\end{equation}
and is an estimator of the free-energy difference (times $-\beta$)
between the dominant boundary conditions and all other boundary
conditions in sample ${\mathcal J}$ at inverse temperature $\beta$.
(Henceforth, we leave ${\mathcal J}$ and $\beta$ implicit). Let
$G_L(\lambda)$ be the cumulative distribution function for $\lambda$,
then it was shown in Ref.~\cite{wang:14} that the function
$1-G_L(\lambda)$ is approximately exponential, which then allows for a
scaling analysis. Define a characteristic $\lambda_{\rm char}(L)$ such
that $1-G(\lambda) = e^{-\lambda/\lambda_{\rm char}}$ and
$1-G(\lambda_{\rm char} \log b) = 1/b$ for any $b$. The value $b$ should
be chosen such that $\lambda_{\rm char}$ is obtained from the tail of
the distribution but not so far into the tail where the statistics are
poor. For $T=0.5$ (dimensionless units), we choose $b = 10$. We have
verified that the distribution functions for different linear system
sizes $L$ collapse well onto the same curve after being scaled by
$\lambda_{\rm char}(L)$.  Standard spin stiffness scaling then gives
\begin{equation}
\label{eq:thetalam}
\lambda_{\rm char} \sim L^{\theta} .
\end{equation} 

A key element of the analysis of both bond chaos and temperature chaos
is the identification and analysis of boundary condition {\em
crossings}, which occur at values of ${\mathcal J}$ and $\beta$ where
there are two boundary conditions $i$ and $j$ having the same fraction
in the population, $p_i = p_j$. Let $|\Delta E|$ be the absolute value
of the energy difference at a boundary condition crossing. Then the
scaling of $|\Delta E|$ with the system size $L$ yields the exponent
controlling the domain-wall fractal dimension ($\ds$) according to,
\begin{equation}
\langle |\Delta E| \rangle \sim L^{\ds} .
\end{equation}
Here, the average is over all crossing in a given range of temperature
or bond configuration and above a threshold $p_c$, such that at the
crossing $p_i=p_j > p_c$. We have also used the median of $|\Delta E|$
rather than the mean.

Boundary condition crossings are manifestations of chaos. The scaling of
their number with the system size gives access to the chaos exponent
$\zeta$. Let $N_C$ be the number of {\em dominant crossings} in some
range of either $\beta$ or $\mathcal{J}$. At a dominant crossing, the
two boundary conditions exchange dominance and on either side of the
crossing, one of the two boundary conditions is dominant. Dominant
crossings, rather than all crossings above an arbitrary threshold, are
used to reduce finite-size effects. From $N_C$ within some range of
$\beta$ or $\mathcal{J}$, we compute $\zeta$ from,
\begin{equation}
N_C \sim L^{\zeta} .
\end{equation}
The relationship between the exponents presented in
Eq.~\eqref{eq:relation}, as well as the relative strength of bond and
temperature chaos (which corresponds to the ratio of the scaled density
of the number of dominant boundary condition crossings $N_C$), are
examined and discussed in Sec.~\ref{results}.

\subsection{Simulation details}
\label{sec:sim}

We start by discussing how to simulate bond chaos for a single disorder
realization. In the reduced Hamiltonian representation, we can either
change the inverse temperature $\beta$ or perturb the spin-spin
interactions ${\mathcal J}$.  We use the following procedure to change
the interactions: For each disorder realization ${\mathcal J}_0$, we
choose an independent perturbation ${\mathcal J}'$ and change the
original bonds as
\begin{equation}
{\mathcal J} =\dfrac{{\mathcal J}_0+c{\mathcal J}'}{\sqrt{1+c^2}},
\end{equation}
where $c \in[0,0.1]$ is a small number. In this manner, for each value
of the perturbation strength $c$, a Gaussian disorder distribution is
preserved \cite{neynifle:97,neynifle:98,krzakala:05,katzgraber:07}.

From now on, within the reduced Hamiltonian representation, we vary the
parameters $\beta$ and $c$ in the $(\beta,c)$ plane. We start the
simulation first with fixed $c$ from $\beta=0$ down to $\beta=2$. This
takes the system from the paramagnetic to the spin-glass phase.
Following this anneal in temperature, we fix $\beta=2$ and change the
bond perturbation strength $c$ in the interval $[0,0.1]$ to induce
chaotic effects. To double-check our results, we have chosen the
interactions of the unperturbed system, ${\mathcal J}_0$, from the study
of temperature chaos in Ref.~\cite{wang:15a}.  The perturbed
interactions ${\mathcal J}'$ were chosen independently. We first do a
temperature anneal of the system from $\beta=0$ to $\beta=2$ at $c=0.1$
fixed, and then we change $c$ from $0.1$ to $0$. In this way, the final
interaction configuration is the same as ${\mathcal J}_0$, which allows
us to compare the results directly to the ones from our temperature
chaos study in Ref.~\cite{wang:15a}.  It is of paramount importance to
verify after both simulation paths that the weights of each boundary
conditions $\{p_i\}$ agree.  In our simulations, we require the family
entropy $S_f \geq \log(100)$ (where $S_f$ is a measure of equilibration
discussed in Ref.~\cite{wang:15e}) for each path as well as
$\max\{|p_i-p_i^\prime|\} \leq 0.05$, where $\{p_i\}$ and
$\{p_i^\prime\}$ are the weights of each boundary conditions from the
two distinct paths, respectively, for each sample.

\begin{figure}[htb]
\begin{center}
\includegraphics[scale=0.7]{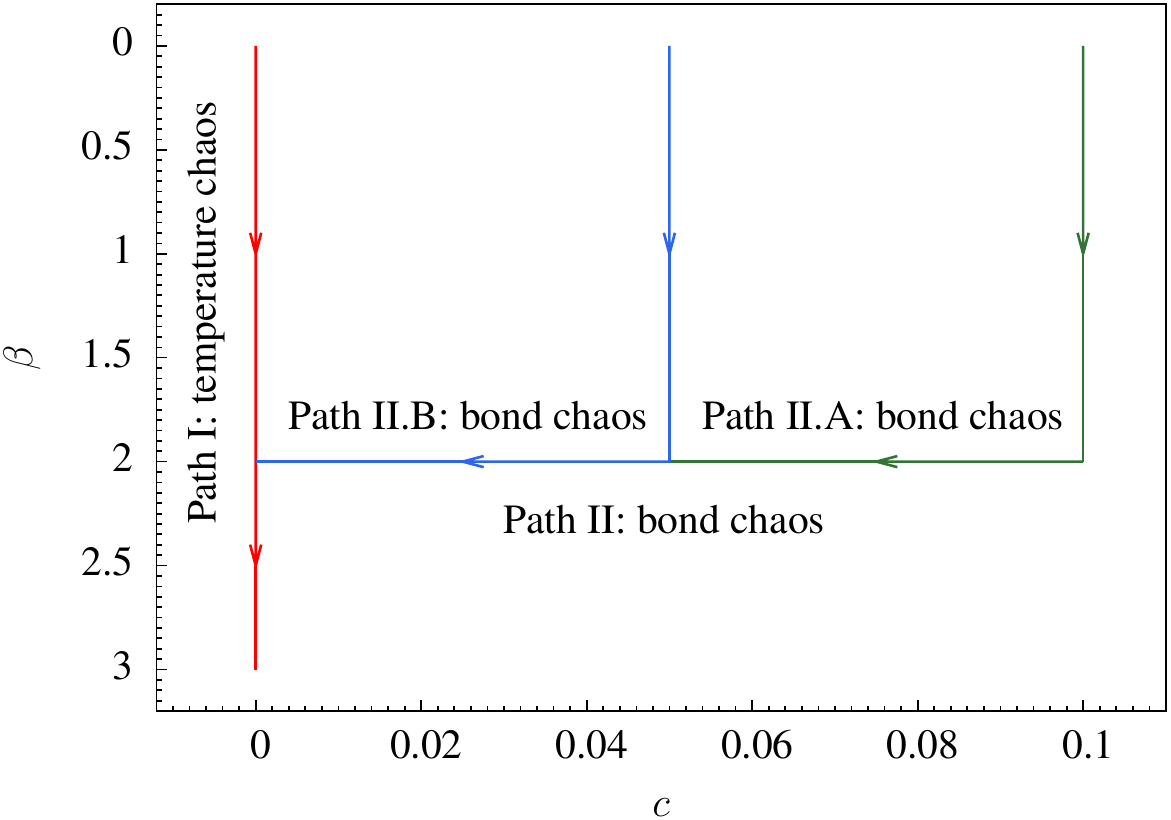}
\caption{(Color online) 
Schematic simulation paths. In all cases, a quench from $\beta = 0$ to
$\beta = 2$ is first performed. For the bond chaos case, an additional
set of simulations for $\beta = 2$ along the $c$-axis from $c = 0.1$ to
$c = 0$ is performed.  Path I (red) represents a simulation to probe
temperature chaos. Path II represents a bond chaos simulation. If the
sample is computationally difficult, then path II is split into paths
II.A (green) and II.B (blue) that are each run independently.}
\label{path}
\end{center}
\end{figure}

Computationally hard samples that do not fulfill the two equilibration
criteria were either rerun with a larger population size, or by breaking
the bond-chaos $c$ path into two segments, where each segment separately
is considerably easier to equilibrate. Figure \ref{path} shows how the
$c$ path is split into two pieces (path II.A and II.B). Measurements
along path II.B require an additional population annealing run starting
at $\beta=0$ and $c=0.05$. Whenever runs are combined, we test the
family entropy of each run, as well as the matching of boundary
conditions between the different runs.

We have carried out a large-scale simulation of bond chaos for linear
system sizes $L=4$, $6$, $8$, and $10$, with $2001$ samples for each
system size. The details of the simulation parameters are summarized in
Table.~\ref{para}.

\begin{table}
\caption{
Parameters of bond-chaos simulations using generalized PA.  $L$ is the
linear system size, $R_0$ is the standard number of replicas, $T_0 =
1/\beta_0$ is the lowest temperature simulated, $N_T$ is the number of
temperature steps (evenly spaced in $\beta$) in the annealing schedule,
$N_b$ is the number of disorder steps (evenly spaced in $c$) in
the annealing schedule, and $n$ is the number of disorder realizations
studied.
\label{para}
}
\begin{tabular*}{\columnwidth}{@{\extracolsep{\fill}} l c c c c c}
\hline
\hline
$L$  & $R_0$  & $T_0$ & $N_T$ & $N_b$ & $n$ \\
\hline
$4$  & $5\times10^4$ & $0.5$  & $101$ & $51$ & $2001$ \\
$6$  & $2\times10^5$ & $0.5$  & $101$ & $51$ & $2001$ \\
$8$  & $5\times10^5$ & $0.5$  & $201$ & $101$ & $2001$ \\
$10$ & $2\times10^6$ & $0.5$  & $301$ & $101$ & $2001$ \\
\hline
\hline
\end{tabular*}
\end{table} 

\section{Results}
\label{results}

\subsection{Scaling properties of bond chaos}

\begin{figure}[htb]
\begin{center}
\includegraphics[scale=0.7]{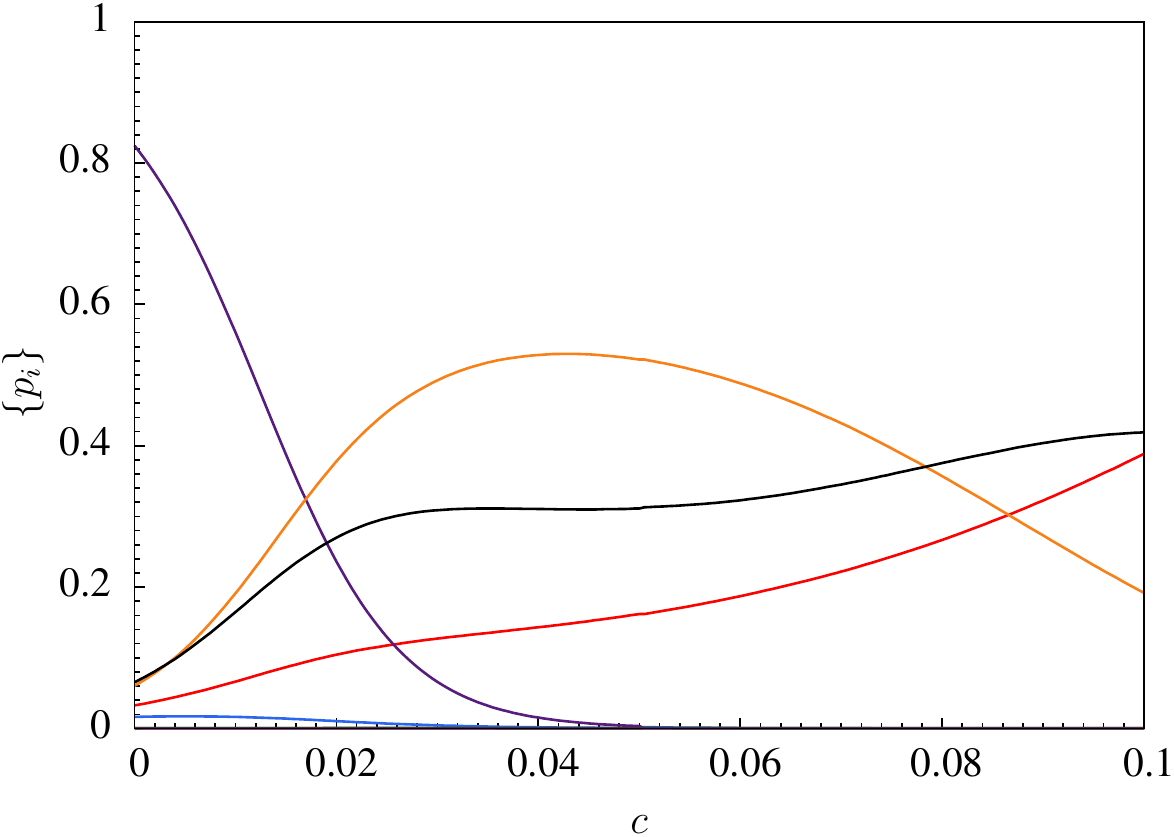}
\caption{(Color online) 
Evolution of the relative weights of each boundary conditions,
$\{p_i\}$, for a chaotic sample of linear system size $L=8$ and
$\beta=2$. Different boundary conditions show crossings, i.e., chaotic
events.  Three of the eight boundary conditions have probabilities too
small to be seen on the plot.
}
\label{cs}
\end{center}
\end{figure}

\begin{figure}[htb]
\begin{center}
\includegraphics[scale=0.7]{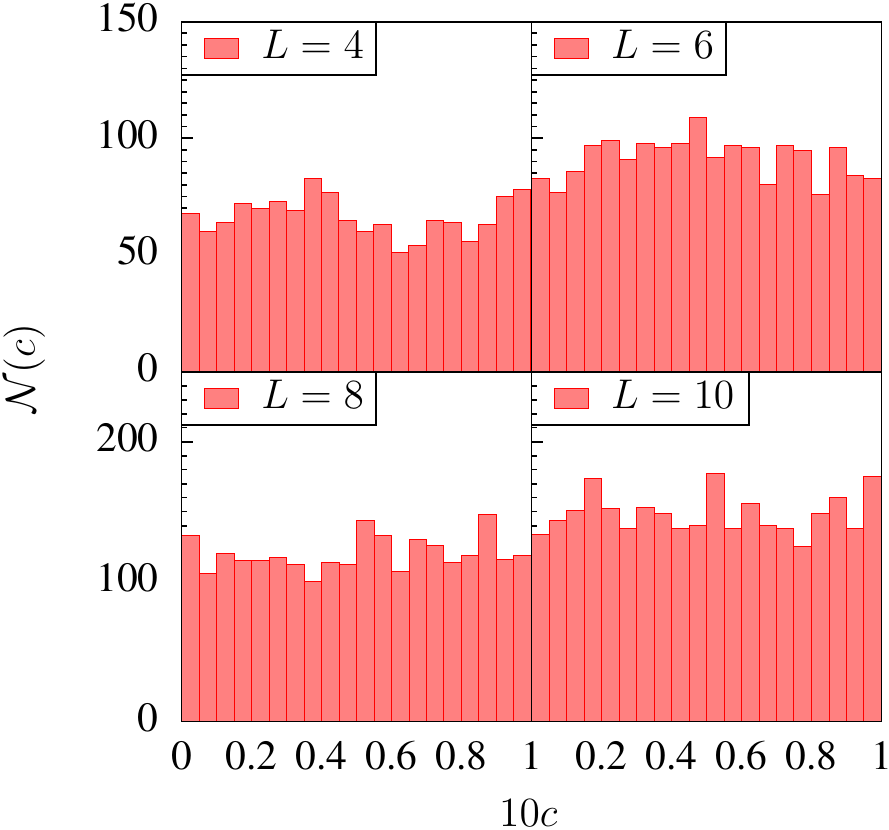}
\caption{(Color online)
Distribution of all crossings above a cutoff of $p_c=0.1$ as a function
of $c$. The distributions are relatively uniform compared to the
distributions of temperature chaos \cite{wang:15a} which are exponential.
Note that the horizontal axis has been multiplied by 10 for better
viewing.}
\label{nc}
\end{center}
\end{figure}

The boundary condition probabilities of a sample of linear size $L=8$ is
shown in Fig.~\ref{cs}, displaying chaotic behavior via several boundary
condition crossings.  In this figure and in the analysis of the energy
differences, we have registered crossings above a threshold of $p_c=0.1$
when two boundary conditions cross as a function of $c$. Figure~\ref{nc}
shows a histogram of the distribution (i.e., number density) of
crossings as a function of $c$, which is relatively flat, as expected.
On the other hand, the distribution of the crossings is approximately
exponential as a function of $\beta$ when chaotic effects are induced by
thermal changes \cite{wang:15a}.

\begin{figure}[htb]
\begin{center}
\includegraphics[scale=0.7]{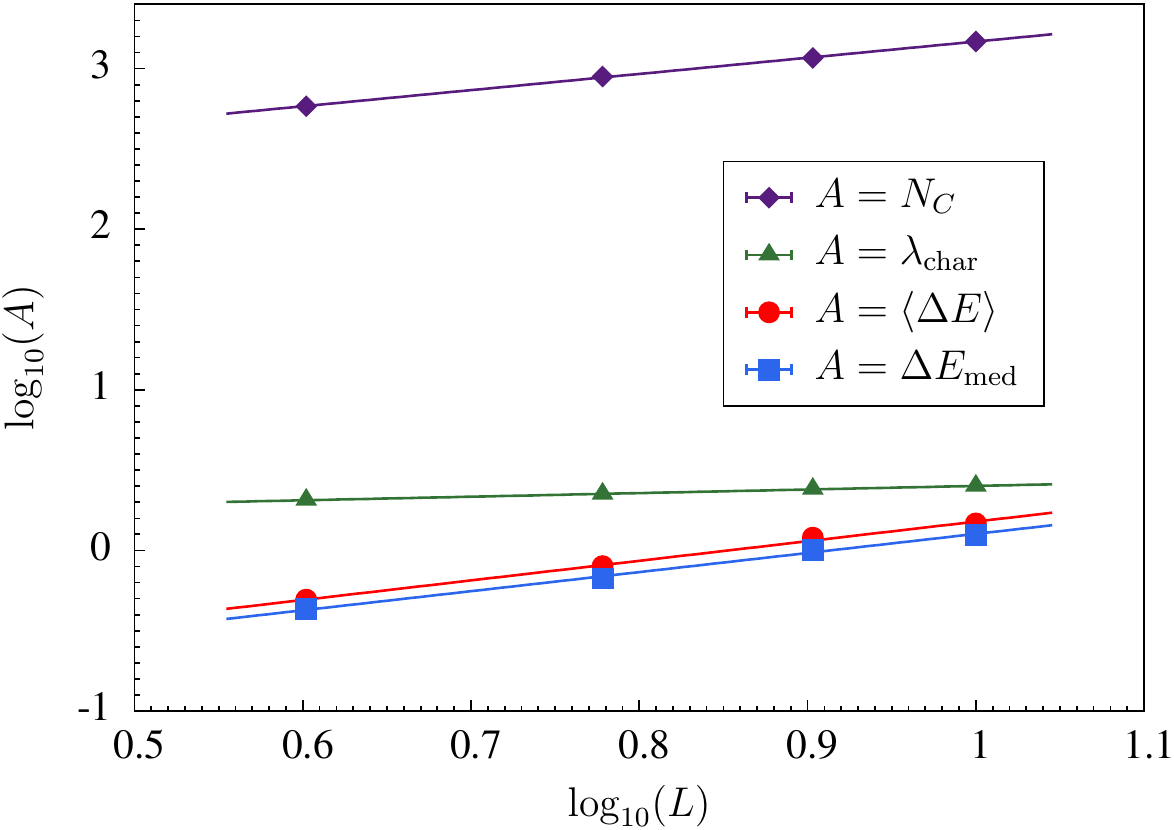}
\caption{(Color online)
Scaling of the measured quantities as a function of system size. The
log-log plot clearly shows that the different quantities are well fit
with a power law. The slope of the number of dominant crossings scales
as $N_C \sim L^\zeta$, the energy difference at all registered boundary
condition crossings above $p_c=0.1$ scales as $\Delta E \sim L^{\ds}$, and
$\lambda_{\rm char} \sim L^\theta$.  Note that for the energy difference
at all registered boundary condition crossings both the median $\Delta
E_{\rm med}$ and average $\langle \Delta E \rangle$ have the same slope
within error bars.  Error bars are smaller than the symbols.}
\label{scaling}
\end{center}
\end{figure}

The system-size scaling of the number of dominant crossings $N_C$,
sample stiffness $\lambda_{\rm{char}}$, mean and median of the energy
cost at boundary condition crossings $\langle | \Delta E | \rangle$ and
$|\Delta E|_{\rm{med}}$, respectively, are all shown in
Fig.~\ref{scaling}.  We have used data at both $(\beta=2,c=0)$ and
$(\beta=2,c=0.1)$ after a temperature anneal for the scaling of
$\lambda_{\rm{char}}$ to improve statistics for the measurement of
$\theta$. The different exponents can be extracted by linear fits to the
data. Our estimates are
\begin{eqnarray}
\theta &=& 0.22(3) \; ,					\\
\ds    &=& 1.22(3) \;\;\;\;\;\; {\rm (mean)} \; ,   	\\
\ds    &=& 1.19(3) \;\;\;\;\;\; {\rm (median)} \; ,	\\
\zeta  &=& 1.01(4)\; .
\end{eqnarray}
Note that $\ds-\theta = 1.00(4)$ (mean) and $\ds-\theta = 0.97(4)$
(median), are in good agreement with $\zeta$. These estimates of the
exponents are also in agreement with the results obtained from
temperature chaos \cite{wang:15a}, showing that temperature chaos and
bond chaos indeed share the same set of scaling exponents.

\subsection{Comparison of temperature and bond chaos}
\label{compare}

\begin{figure}[htb]
\begin{center}
\includegraphics[scale=0.7]{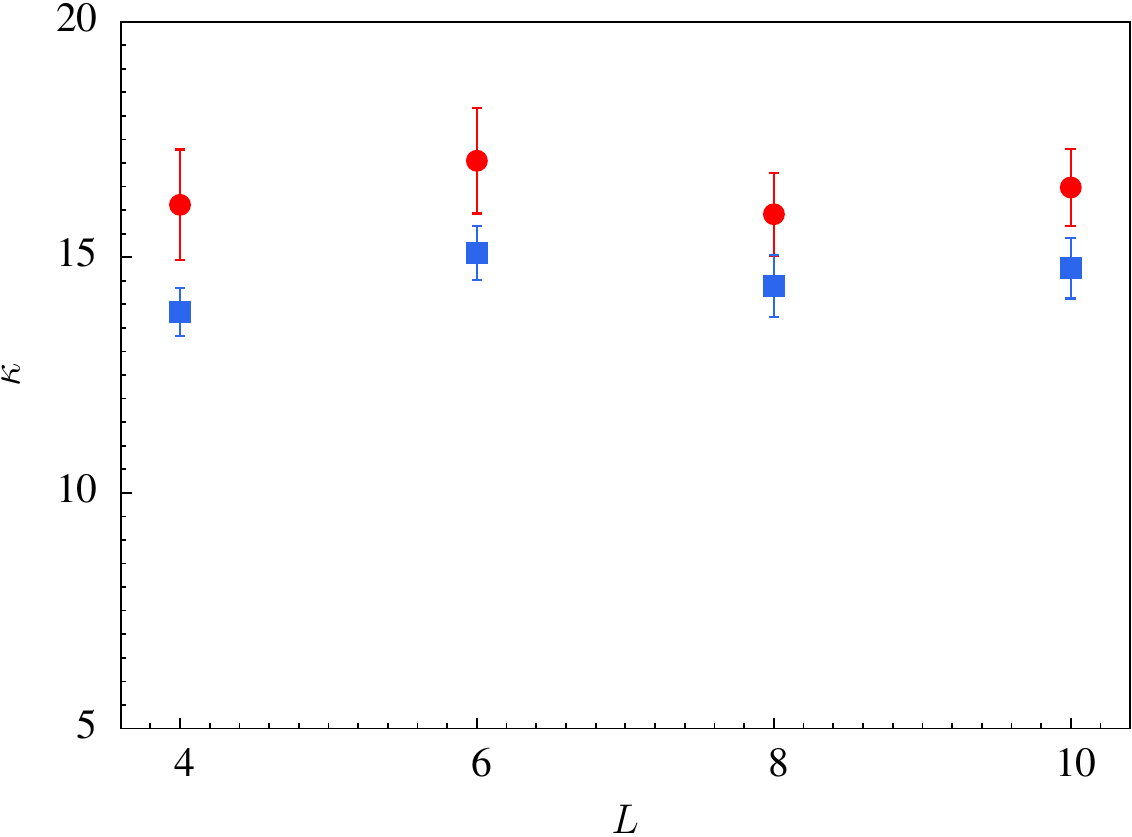}
\caption{(Color online)
Relative strength between bond and temperature chaos, $\kappa$, as a
function of system size $L$ computed from Eq.~\eqref{E1} (red circles)
and Eq.~\eqref{E2} (blue squares). Using droplet scaling arguments one
can show that $\kappa$ does not depend on the system size.}
\label{eta}
\end{center}
\end{figure}

Next, we compare the relative strength of temperature chaos and bond
chaos. A natural quantity to compare is the number of dominant crossings
$N_C$. However, due to the different types of perturbations (temperature
$\beta$ vs interactions $c$), we need to compare fairly the density of
crossing with respect to $\beta$ and $c$. Let $\Delta \beta$ be the
range of inverse temperatures in the analysis of temperature chaos and
$\Delta c$ (here $\Delta c=0.1$) the range of modification of the bond
configuration. The relative strength of the perturbation of temperature
chaos to bond chaos seen from the reduced Hamiltonian is
\begin{equation}
\frac{\Delta \beta}{\beta \Delta c} . 
\end{equation}
The distribution of crossings of bond chaos is approximately uniformly
distributed, therefore, the density of crossings for bond chaos (BC) is
simply given by
\begin{equation}
\rho^{\rm BC}  = \frac{{N_C}^{\rm TC}}{\beta \Delta c} .
\label{eq:aa} 
\end{equation}
The distribution of crossing for temperature chaos is more complicated
and is approximately exponential in the range $\beta \in [\beta_{\rm
min},\beta_{\rm max}] = [1.5,3]$ for all system sizes studied in
Ref.~\cite{wang:15a}. An exponential fit of the form
\begin{equation}
{\mathcal N}(\beta) = 
\frac{a e^{-a \beta}}{e^{-\beta_{\rm min} a} - e^{-\beta_{\rm max} a}} , 
\end{equation}
with $\beta \in [1.5,3]$ yields $a \approx 1.12$. This suggests that the
density of crossing distributions at $\beta=2$ is approximately $1.18$
times larger than that of the averaged density in the whole temperature
range $[1.5,3]$. We note that the ratio depends only very weakly on $a$.
The corresponding density of crossings for temperature chaos (TC) is
therefore given by
\begin{equation}
\rho^{\rm TC} = \frac{1.18{N_C}^{\rm TC}}{\Delta \beta} .
\label{eq:bb}
\end{equation}
Using $\Delta c= 0.1$, $\beta=2$, $\Delta \beta=1.5$, and $N_C$ for the
whole perturbation range of both types of chaos in Eqs.~\eqref{eq:aa}
and \eqref{eq:bb}, we can define a quantity $\kappa$ that quantifies the
relative strength of bond chaos and temperature chaos as
\begin{eqnarray}
\kappa &=&       \frac{\rho^{\rm BC}}{\rho^{\rm TC}} \nonumber \\
       &\approx& 6.34 \frac{{N_C}^{\rm BC}}{{N_C}^{\rm TC}},
\label{E1}
\end{eqnarray}
where ${N_C}^{\rm BC}$ and ${N_C}^{\rm TC}$ are the total number of
dominant boundary condition crossings of bond chaos and temperature
chaos, respectively. A plot of $\kappa$ as a function of the linear
system size $L$ is shown in Fig.~\ref{eta} (red circles). Note that
$\kappa$ is almost a constant function of $L$, as expected from the
scaling properties of $N_C$.  Averaging over the studied system sizes
$L$, we find that
\begin{equation}
\kappa \approx 16(1) . 
\end{equation}
Therefore, bond chaos is more than one order of magnitude stronger than
temperature chaos at low temperatures ($\beta=2$). It is interesting
that our result is very close to the value of $17.5$ quoted in
Ref.~\cite{sasaki:05} even though the two values are computed using
fundamentally different methods and models.  Note that the difference of
the threshold to register crossings does not affect the number of
dominate boundary condition crossings and therefore also $\kappa$, as in
both cases, a dominant boundary condition crossing cannot occur below
the chosen threshold, $p_c=0.1$ for bond chaos and $p_c=0.05$ for
temperature chaos.

We propose a simple physical interpretation for $\kappa$. At first
sight, one might expect that the scale of $\Delta E$ and $\Delta(TS)$
might be relevant to explain $\kappa$. However, while we find this is
indeed a factor---especially at low temperatures---this is not
sufficient. We find that the strength of \textit{responses} of the
quantities with respect to $c$ and $\beta$ are more relevant, as chaos
is a dynamical process. To this effect, we use an alternate definition
of the relative strength $\kappa$, namely
\begin{eqnarray}
\kappa &=& \left\langle
		\frac{\partial |\Delta E|}{\beta \partial c} 
	   \right\rangle 
      \Big{/} 
	   \left\langle
		\frac{\partial |\Delta(TS)|}{\partial \beta} 
	   \right\rangle ,
\label{E2}
\end{eqnarray}
where the derivative is evaluated using finite element methods at the
same temperature $\beta=2$. This theoretical prediction of $\kappa$ is
also shown in Fig.~\ref{eta} (blue squares). Note that Eq.~\eqref{E2}
also explains why $\kappa$ does not depend on the system size from the
same scaling properties of $\Delta E$ and $\Delta(TS)$.  The predictions
are reasonably close, showing that bond chaos is indeed energy driven,
while temperature chaos is entropy driven.

We know that the number of dominant crossings $N_C$ is a growing
function of system size $L$. It is a natural scenario that the fraction
of instances with dominant crossings also grows with $L$, along with the
mean number. The fraction of samples with dominant crossings for both
temperature chaos and bond chaos are shown in Table.~\ref{frac1}.

\begin{table}
\caption{
Fraction of instances with dominant crossings as a function linear
system size $L$ for temperature chaos (TC) and bond chaos (BC). Note
that the fraction of instances with dominant crossings grows with $L$,
along with the mean number. \label{frac1}
}
\begin{tabular*}{\columnwidth}{@{\extracolsep{\fill}} l c c c c}
\hline
\hline
$L$  & $4$  & $6$ & $8$ & $10$ \\
\hline
TC  & 0.11(1) & 0.16(1)  & 0.21(1) & 0.26(1) \\
BC  & 0.27(1) & 0.39(1)  & 0.48(1) & 0.56(1) \\
\hline
\hline
\end{tabular*}
\end{table}

\begin{figure}[htb]
\begin{center}
\includegraphics[scale=0.7]{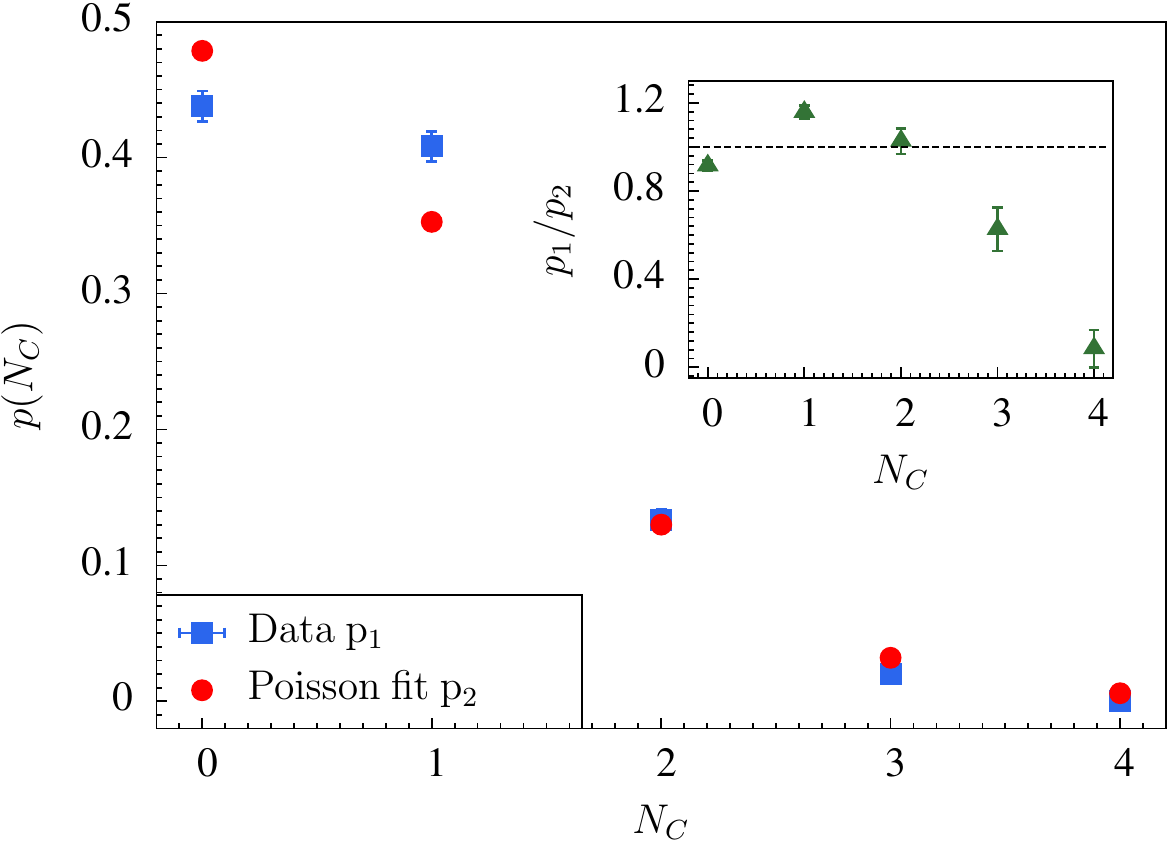}
\caption{(Color online)
Representative distribution of the number of dominant crossings $N_C$
for $L=10$ of bond chaos, along with a Poisson fit. The ratio of the
weights of the distribution to the Poisson fit is shown in the inset.
One can see that the number of dominant crossings is still dominant by
samples with small number of crossings of the majority of instances.
Qualitatively similar results have been obtained for temperature chaos,
as well as all system sizes studied.
}
\label{DNC}
\end{center}
\end{figure}

It has been argued that temperature chaos in spin glasses is dominated
by rare events~\cite{fernandez:13}. By looking at the distribution of
the number of dominant crossing for each disorder realization one can
study this hypothesis in the context of both bond and temperature chaos.
Figure~\ref{DNC} shows the probability with respect to disorder
realizations of having $N_C$ dominant crossings as a function of $N_C$
for the case of bond chaos with $L=10$. The Poisson distribution with
the same mean is also shown in Fig.~\ref{DNC}. The inset shows the ratio
of both distributions.  There are systematic deviations between the data
and the Poisson distribution, specifically there are fewer $N_C=0$
samples than the Poisson distribution and more $N_C=1$ samples. More
importantly, the probability of getting large values of $N_C$ is less
than the Poisson prediction.  We find qualitatively similar results for
both temperature and bond chaos and for all sizes studied.  Therefore,
in contrast to the results of Ref.~\cite{fernandez:13}, our measure of
chaos based on boundary condition crossings is {\em not} described by an
extreme value distribution and is not dominated by rare events.

It has recently been established that temperature chaos is associated
with computational hardness
\cite{fernandez:13,wang:15a,zhu:16,martin-mayor:15}.  We note that the
same can be stated for bond chaos. In Sec.~\ref{sec:sim} we divided
disorder realizations into two groups: hard to simulate and easy to
simulate. We take a closer look at the average number of dominant
crossings in each class. For $L=8$, approximately $13\%$ of the
instances are typically hard, and the mean number of dominant crossings
are $0.92(4)$ and $0.53(2)$ for computationally ``hard'' and ``easy''
instances, respectively. For $L=10$, approximately $47\%$ of the
instances are hard, and the mean number of dominant crossings are
$1.02(3)$ and $0.49(2)$ for hard and easy instances, respectively.

This suggests that the difficulty of transforming an equilibrium state
for one set of bonds to another set of bonds is strongly correlated with
bond chaos along the path connecting the two bond configurations.  This
is not unexpected because a crossing indicates that configurations
(including both spin and boundary condition) that were important for one
set of bonds are no longer as important when the bonds are modified. The
fact that both temperature and bond chaos introduce computational
hardness, suggests the possibility of optimizing population annealing
Monte Carlo for simulating a fixed bond configuration by choosing curved
paths in the temperature--disorder space that minimizes chaos. A
simulation of disorder realization $\mathcal{J}$ might begin with
disorder realization $\mathcal{J}'$ and involve annealing in both
temperature and bond strength. This idea remains to be explored and
might also provide an avenue to overcome gaps in the energy spectrum for
quantum annealing simulations.

\section{Conclusions}
\label{cc}

In this work, we have studied bond chaos using thermal boundary
conditions via a generalization of population annealing Monte Carlo. We
provided a simple explanation as to why temperature chaos and bond chaos
share the same set of scaling exponents within the framework of the
droplet/scaling picture. We also show quantitatively that bond chaos is
approximately one order of magnitude stronger than temperature chaos. A
simple physical picture is proposed that explains the relative strength,
identifying that bond chaos is energy driven, whereas temperature chaos
is entropy driven. As such, the surface of excitations plays a key role
in these phenomena.  Our work on temperature and bond chaos also
establishes the validity of the use of thermal boundary conditions
to study chaotic phenomena in disordered systems.

We use this opportunity to emphasize that the fact that bond chaos is so
much stronger than temperature chaos should be a source of concern in
analog computing machines, such as the D-Wave Systems Inc.~DW2X.  While
slight temperature variations of the device might only affect the
encoded problem slightly, small perturbations of the couplers might
change the encoded Hamiltonian and yield erroneous results.

We intend to apply the method used here to other paradigmatic spin-glass
systems, such as the four-dimensional Edwards-Anderson Ising spin glass.
A more challenging and intriguing problem is to investigate field chaos
in the mean-field Sherrington-Kirkpatrick model
\cite{sherrington:75}---which also has a spin-glass phase in an external
magnetic field---using a generalized definition of boundary conditions.

\acknowledgments 

J.M.~acknowledges support from National Science Foundation (Grants
No.~DMR-1208046 and DMR-1507506).  H.G.K.~acknowledges support from the
National Science Foundation (Grant No.~DMR-1151387) and would like to
thank Sch{\"o}fferhofer for support during the manuscript preparation.
W.W.~acknowledges support from National Science Foundation (Grants
No.~DMR-1208046 and DMR-1151387). The research of H.G.K.~is based upon
work supported in part by the Office of the Director of National
Intelligence (ODNI), Intelligence Advanced Research Projects Activity
(IARPA), via MIT Lincoln Laboratory Air Force Contract
No.~FA8721-05-C-0002.  The views and conclusions contained herein are
those of the authors and should not be interpreted as necessarily
representing the official policies or endorsements, either expressed or
implied, of ODNI, IARPA, or the U.S.~Government.  The U.S.~Government is
authorized to reproduce and distribute reprints for Governmental purpose
notwithstanding any copyright annotation thereon. We thank Texas A\&M
University for access to their Ada and Curie clusters.

\bibliography{refs}

\end{document}